\documentclass[preprint,preprintnumbers,amsmath,amssymb]{revtex4}
\usepackage{graphicx}
\usepackage{color}
\usepackage{units}

\newcommand{\abs}[1]{\left\vert#1\right\vert}

\newcommand{\ket}[1]{\vert #1 \rangle}

\definecolor{shadecolor}{rgb}
{0.9,0.8,1}

\begin{document}
\title{Quantum Walks in an array of Quantum Dots}
\author{K. Manouchehri}
%\email{kia@physics.uwa.edu.au}
\author{J.B. Wang}
\email{wang@physics.uwa.edu.au} \affiliation{School of Physics, The
University of Western Australia}
\date{\today}

\begin{abstract}
    Quantum random walks are shown to have non-intuitive
    dynamics, which makes them an attractive area of study for
    devising quantum algorithms for well-known classical problems as well
    as those arising in the field of quantum computing. In this work
    we propose a novel scheme for the physical implementation of a
    discrete-time quantum random walk using laser excitations of the
    electronic states of an array of quantum dots. These dots represent the discrete nodes of the walk, while transitions between the energy levels inside each dot correspond to the required coin operation and stimulated Raman adiabatic passage (STIRAP) processes are employed to induce the steps of the walk. The quantum dot design is tailored in such a way as to enable selective coupling of the energy levels. Our simulation results show a close agreement with the ideal quantum walk distribution as well as modest robustness towards noise disturbance.
\end{abstract}


\maketitle

\section{Introduction}

Quantum random walks represent a generalized version of the well
known classical random walk, which can be elegantly described using
quantum information processing terminology \cite{Aharonov1993}.
Despite their apparent connection however, dynamics of quantum
random walks are often non-intuitive and deviate significantly from
those of their classical counterparts \cite{Farhi1998}. Among the
differences, the faster mixing and hitting times of quantum random
walks are particularly noteworthy, making them an attractive area of
study for devising efficient quantum algorithms, including those
pertaining to connectivity and graph theory \cite{Kempe2003,
Farhi1998, Childs2003}, as well as quantum search algorithms
\cite{Shenvi2003, Childs2004}.

There are two broad classes of quantum random walks, namely the
discrete- and continuous-time quantum random walks, which have
independently emerged out of the study of unrelated physical
problems. Despite their fundamentally different quantum dynamics
however, both families of walks share similar and characteristic
propagation behavior \cite{Kempe2003, Konno2005, Patel2005}.
Strauch's recent work \cite{Strauch2006} is the latest in a line of
theoretical efforts to establishing a formal connection between the
discrete and continuous-time quantum random walks, in a manner
similar to their classical counterparts.

There have been several proposals for implementing quantum random
walks using a variety of physical systems including Nuclear Magnetic
Resonance \cite{Du2003, Ryan2004}, cavity QED \cite{Aharonov1993,
Agarwal2005, Sanders2003, Di2004}, ion traps \cite{Travaglione2002},
classical optics \cite{Zhao2002, Jeong2004, Do2005, Knight2003-1,
Knight2003-2, Banuls2006, Francisco2006}, quantum optics
\cite{Zou2006, Zhang2007}, optical multiports \cite{Hillery2003,
Feldman2004, Kovsik2005}, optical lattice and microtraps
\cite{Dur2002, Chandrashekar2006, Joo2006, Cote2006, Eckert2005} as
well as quantum dots \cite{Solenov2006}.

In this paper we introduce another proposal for implementing the
\emph{discrete-time} or coined quantum walk on a line using a series
of stimulated Raman adiabatic passage (STIRAP) operations \cite{Bergmann1998, Kral2001, Hohenester2000} on a single electron trapped in an array of quantum dots. An important advantage of our proposal is that it relies on well
established and generally accessible experimental techniques which
result in both high fidelity operations as well as a relative ease
of scalability. To the best of our knowledge, the proposal of
Solenov and Fedichkin \cite{Solenov2006} is the only other
implementation to date which employs quantum dots, but unlike our
scheme, it pertains to a \emph{continuous-time} quantum walk on a
circle.

In what follows we present a brief overview of the coined quantum
random walk (Sec. \ref{section.qrw-overview}) and describe our
proposal for the design of the quantum dot array and the sequence of
required STIRAP operations (Sec. \ref{section.proposal}). We then
present a numerical simulation of the system's evolution (Sec.
\ref{section.results}), including the effect of imperfect STIRAP
operations. In the appendix we also demonstrate an efficient
numerical technique for the optimization of STIRAP pulse parameters.

\section{Coined Quantum Random Walk} \label{section.qrw-overview}

A one-dimensional quantum random walk consists of a walker hopping
between $N$ nodes or quantum states $\ket{i}$ ($-N/2-1<i<N/2$)
assembled in a line. In the coined quantum walk, each state
$\ket{i}$ further consists of two sub-levels or coin states labeled
as $\ket{\uparrow, i}$ and $\ket{\downarrow, i}$. Unlike the
classical case, the quantum walker has a complex valued distribution
$\psi$ over all the states, which remains undetected throughout the
walk. Each step of the walk involves a \emph{coin flip}, defined as
a simultaneous unitary rotation
\begin{equation} \label{eqn.unitary-coin}
    \hat{C}(\theta, \phi_1, \phi_2) = \left(
    \begin{array}{cc}
        \cos(\theta) & \sin(\theta)e^{i\phi_1} \\
        \sin(\theta)e^{i\phi_2} & -\cos(\theta)e^{i(\phi_1+\phi_2)}
    \end{array} \right),
\end{equation}
on the coin states of all nodes, followed by a conditional
translation which shifts the walker in states $\ket{\uparrow, i}$
and $\ket{\downarrow, i}$ to states $\ket{\uparrow, i+1}$ and
$\ket{\downarrow, i-1}$ respectively. Hence for a quantum walker in
an initial state $\psi(0)$, its state after $n$ steps of the walk is
given by $\psi(n)= \hat{U}^n ~\psi(0)$, where
\begin{equation}\label{eqn.evolution-operator}
    \hat{U} = \hat{T}^{\downarrow}_{-1}~\hat{T}^{\uparrow}_{+1}~\hat{C}
\end{equation}
is the overall evolution operator for a single step. A final
probability distribution is determined by collapsing the walker's
wavefunction $\psi(n)$ at the end of the evolution.

In this paper, we implement a modified evolution operator
\begin{equation}
    \widetilde{U} = \hat{T}^{\uparrow}_{+1}~\hat{C}
\end{equation}
which is the same as Eq. $\ref{eqn.evolution-operator}$ up to a
translation and relabeling of states. In other words we can define a
mapping $\hat{M}: \widetilde{U} \longmapsto \hat{U}$ where $\hat{M}=
\hat{T}_{-1} ~ \hat{L}$ first relabels all the nodes according to
$\hat{L}: \ket{i} \longmapsto \ket{2i}$, followed by a translation
$\hat{T}_{-1}$ of the entire wavefunction one node to the left, that
is
\begin{eqnarray}
    \nonumber   \hat{M} ~ \widetilde{U}
    \nonumber   &=& \hat{T}_{-1} ~ \hat{L} ~~ \hat{T}^{\uparrow}_{+1}~\hat{C} \\
    \nonumber   &=& \hat{T}_{-1} ~ ~ \hat{T}^{\uparrow}_{+2}~\hat{C} \\
    \nonumber   &=& \hat{T}^{\downarrow}_{-1} ~ \hat{T}^{\uparrow}_{-1} ~ \hat{T}^{\uparrow}_{+2}~\hat{C} \\
    \nonumber   &=& \hat{T}^{\downarrow}_{-1} ~ \hat{T}^{\uparrow}_{+1}~\hat{C} \\
                &=& \hat{U}.
\end{eqnarray}

\section{Physical Implementation} \label{section.proposal}

\begin{figure}[h]
    \includegraphics[width=11cm, bb=0 0 530 310,clip]{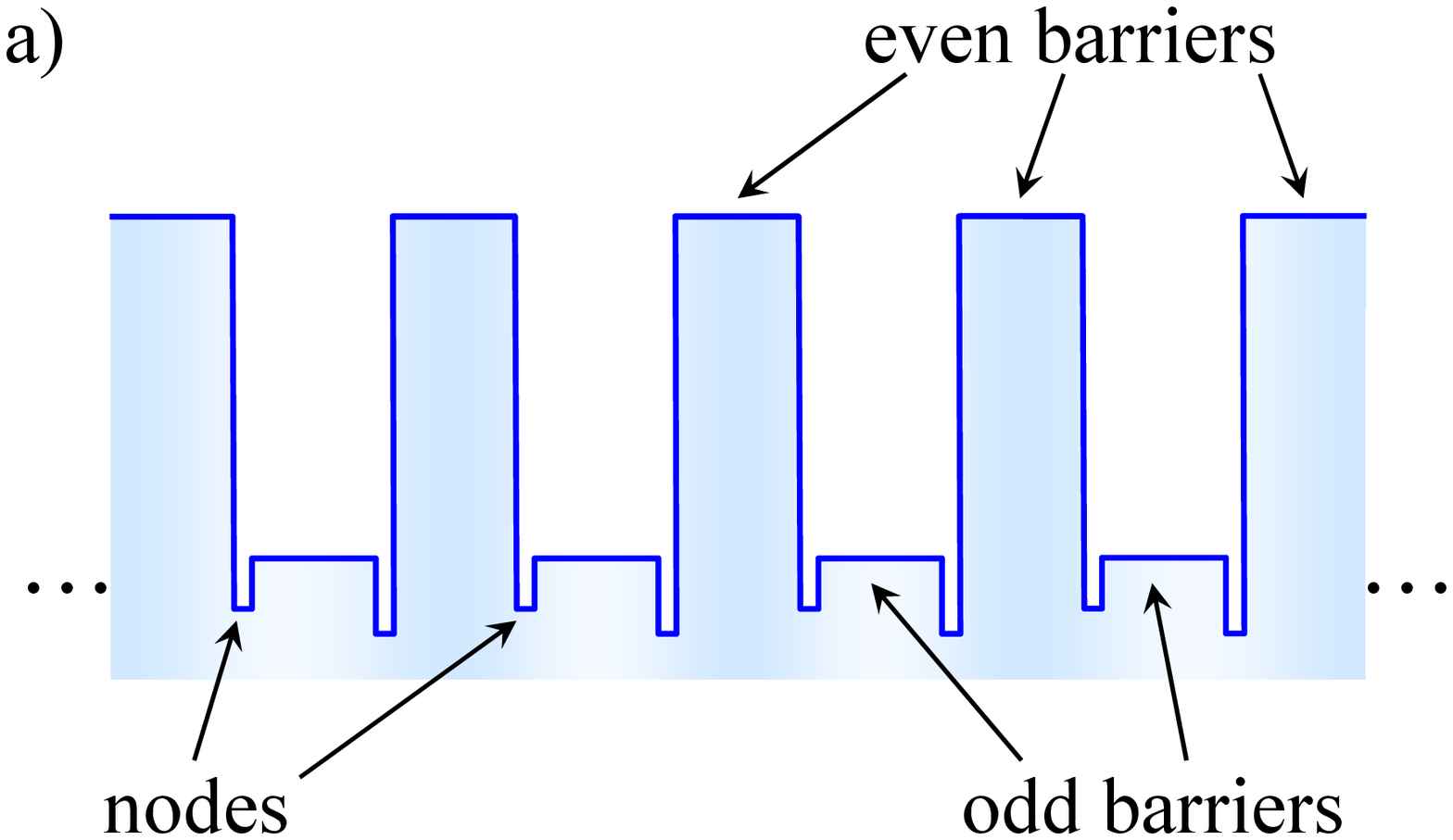}
    \includegraphics[width=11cm, bb=0 0 530 490,clip]{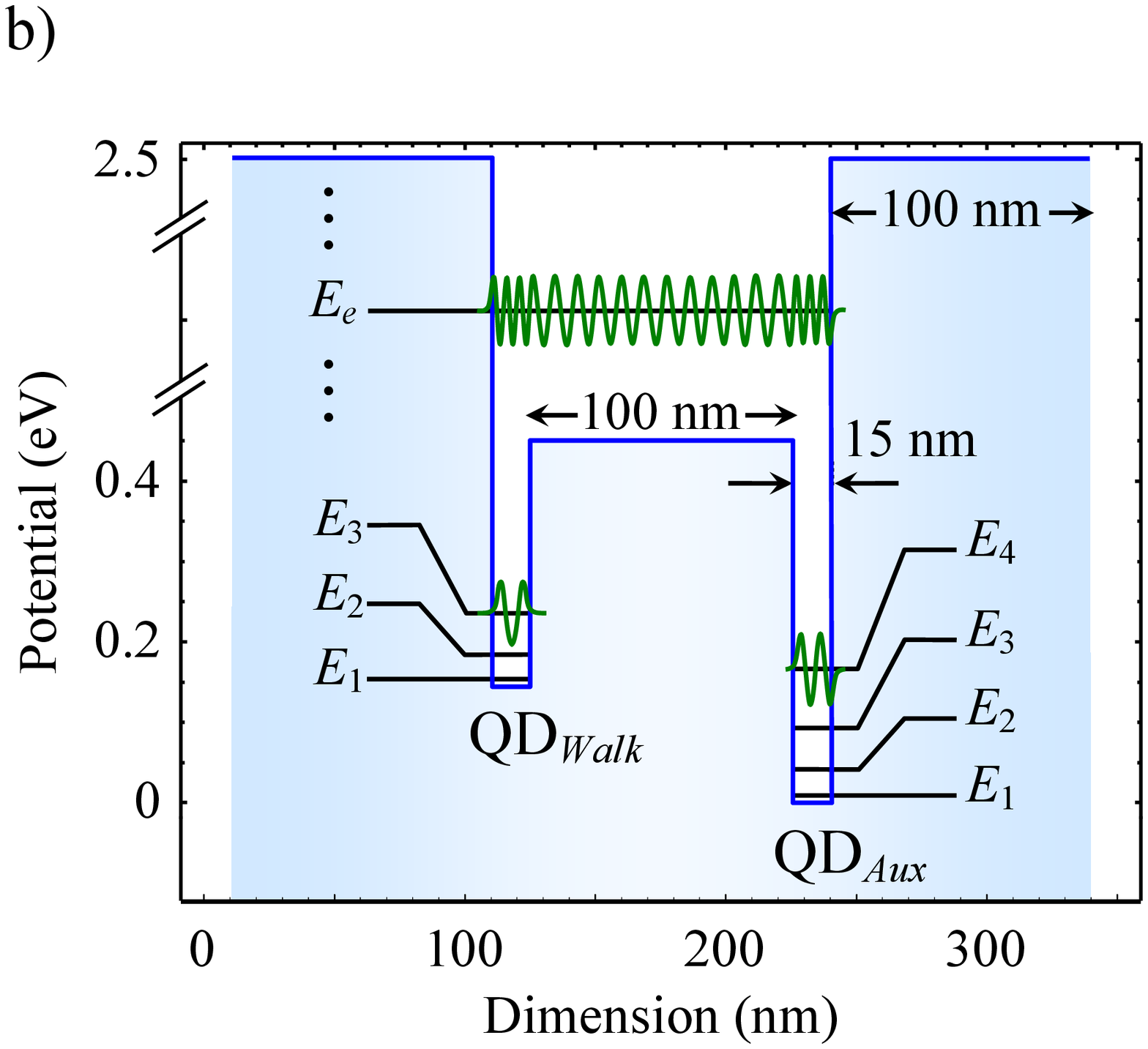}
    \caption{a) An array of quantum dots representing the discrete nodes for a quantum walk on a line.
    b) The electronic structure of a pair of quantum dots $\text{QD}_{Walk}$ and $\text{QD}_{Aux}$.
    For the first few energy eigenstates, the overlap between the electron wavefunction inside the dots is negligible.}
    \label{fig.qdot-node-scheme}
\end{figure}

To implement the quantum walk we use an array of quantum dots,
depicted in Fig. \ref{fig.qdot-node-scheme}a, where all the odd
barriers have been significantly lowered creating pairs of coupled
dots. Figure \ref{fig.qdot-node-scheme}b illustrates a more detailed
structure of the pair, labeled as the walk quantum dot
$\text{QD}_{Walk}$ and the auxiliary quantum dot $\text{QD}_{Aux}$.
The essential feature of this design is that for low energies, the
energy eigenstates of $\text{QD}_{Walk}$ and $\text{QD}_{Aux}$ are,
to a very good approximation, spatially separable and the electron
wavefunction is localized within the dot. For energies above their
joint potential barrier however the two dots share common electronic
states.

\begin{figure}
    \includegraphics[width=8cm, bb=0 0 370 520,clip]{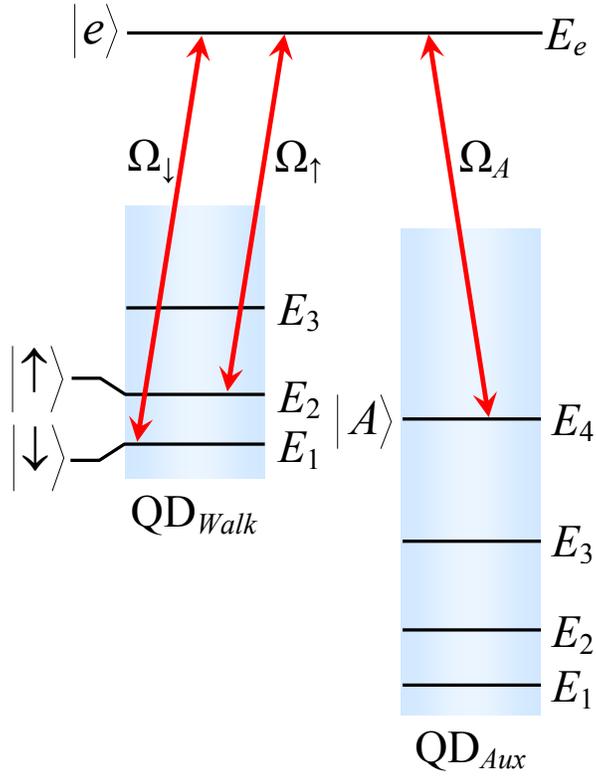}
    \caption{Labeling of the energy levels as coin states
    $\ket{\downarrow}$ and $\ket{\uparrow}$, excited state $\ket{e}$ and auxiliary state $\ket{A}$.
    STIRAP operations between states $\ket{\downarrow}\longleftrightarrow\ket{A}$ and $\ket{\uparrow}\longleftrightarrow\ket{A}$ are facilitates via
    the intermediary state $\ket{e}$, using laser pulses with
    angular frequencies $\Omega_{\downarrow}$, $\Omega_{\uparrow}$ and $\Omega_{A}$.}
    \label{fig.qdot-coupling-stirap}
\end{figure}

The nodes of the quantum walk are mapped to the successive
$\text{QD}_{Walk}$ along the array of quantum dots. As depicted in
Fig. \ref{fig.qdot-coupling-stirap}, the first two energy levels of
$\text{QD}_{Walk}$ encode the coin states $\ket{\downarrow}$ and
$\ket{\uparrow}$ of the walk, the fourth energy level of
$\text{QD}_{Aux}$ represents an auxiliary state $\ket{A}$, $E_e$
represents an excited state $\ket{e}$ well above the joint barrier
between the two dots, and other states remain unoccupied throughout
the walk. The quantum walk itself is represented by the propagation
of a single electron wavefunction through the array of dots using a
series of specially optimized 2- and 3-photon $\Lambda$ STIRAP
operations.

The 2-photon STIRAP is used to perform the translation operation
$\hat{T}^{\uparrow}_{+1}$. Here two laser pulses, pump $P$ and Stoke
$S$, with angular frequencies $\Omega_{\uparrow}$ and $\Omega_{A}$
respectively, couple the dressed states $\ket{\uparrow}$ and
$\ket{A}$ via the intermediate state $\ket{e}$. By tuning the laser
parameters and applying the two pulses in the counter intuitive
sequence, one can achieve coherent population transfer between
states $\ket{\uparrow}$ and $\ket{A}$ with almost perfect fidelity
and without leaving any appreciable papulation residual in state
$\ket{e}$ (See Fig. \ref{fig.2ph-stirap}).

Likewise a pair of 3-photon STIRAP is used to perform the coin
operation $\hat{C}$. First three laser pulses, $P1$ and $P2$ and
$S$, with angular frequencies $\Omega_{\downarrow}$,
$\Omega_{\uparrow}$ and $\Omega_{A}$ respectively, couple the
dressed states $\ket{\downarrow}$, $\ket{\uparrow}$ and $\ket{A}$
via the intermediate state $\ket{e}$. A second 3-photon STIRAP is
then applied in the reverse order. This procedure is shown to be
capable of performing arbitrary rotations on the superposition state
$\alpha \ket{\uparrow} + \beta \ket{\downarrow}$, independently of
the initial amplitudes $\alpha$ and $\beta$ \cite{Kis2002} (See Fig.
\ref{fig.3ph-stirap}). Therefore with careful optimization of laser
parameters we can engineer a variety of coin operators $\hat{C}$.

Figure \ref{fig.qrw-procedure} illustrates how the quantum walk can
be performed. First a pair of 3-photon STIRAP operations perform a
coin rotation $\hat{C}$ simultaneously on all coin states. A
2-photon STIRAP will then transfer all the $\ket{\uparrow,i}$ states
to their corresponding $\ket{A,i}$ state. We then adiabatically
raise all the odd barriers and lower the even barriers, virtually
reversing the paring of the quantum dots. In this new arrangement
every $\text{QD}_{Aux}$ previously associated with the $i$th node is
now paired up with $(i+1)$th $\text{QD}_{Walk}$. Using a second
2-photon STIRAP we can now transfer $\ket{A,i}$ to
$\ket{\uparrow,i+1}$ which completes the implementation of
$\hat{T}^{\uparrow}_{+1}$ operator. The barriers are then returned
to their original setting and the process repeated for additional
steps. The final quantum walk distribution corresponds to the
probability distribution for detecting the electron inside each
$\text{QD}_{Walk}$ in the array of dots.

\begin{figure}
    \includegraphics[width=12cm, bb=0 0 570 480,clip]{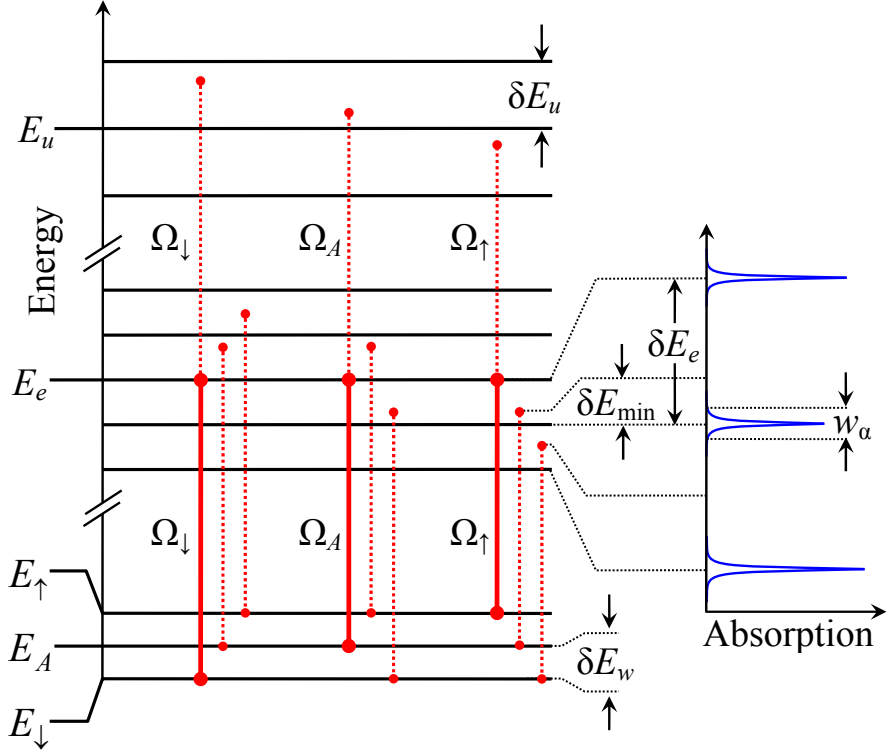}
    \caption{Energy level diagram for a single node corresponding to the quantum dot dimensions presented in Fig.
    \ref{fig.qdot-node-scheme}:
    $E_A \approx 173$ meV, $E_e \approx 1045$ meV and $E_u \approx 1912$
    meV. Energy levels $E_{\uparrow}$ and $E_{\downarrow}$ are
    nearly equidistant from $E_A$ with an energy gap $\delta E_w \approx 15$
    meV. Similarly, the immediate levels above and below $E_e$ and
    $E_u$ are separated by $\delta
    E_e \approx 20$ meV and $\delta E_e \approx 30$ meV respectively.
    The absorption spectrum is assumed to have a line width $w_{\alpha} \lesssim 1$ meV.
    Solid lines represent the coupling between the desired energy
    levels via $\Omega_{\downarrow}$, $\Omega_{\uparrow}$ and
    $\Omega_{A}$ pulses. Dotted lines demonstrate that these frequencies are prevented by the energy structure
    from activating any spurious coupling between other any levels.}
    \label{fig.qdot-selective-coupling}
\end{figure}

\begin{figure}
    \includegraphics[width=8cm, bb=0 0 545 240,clip]{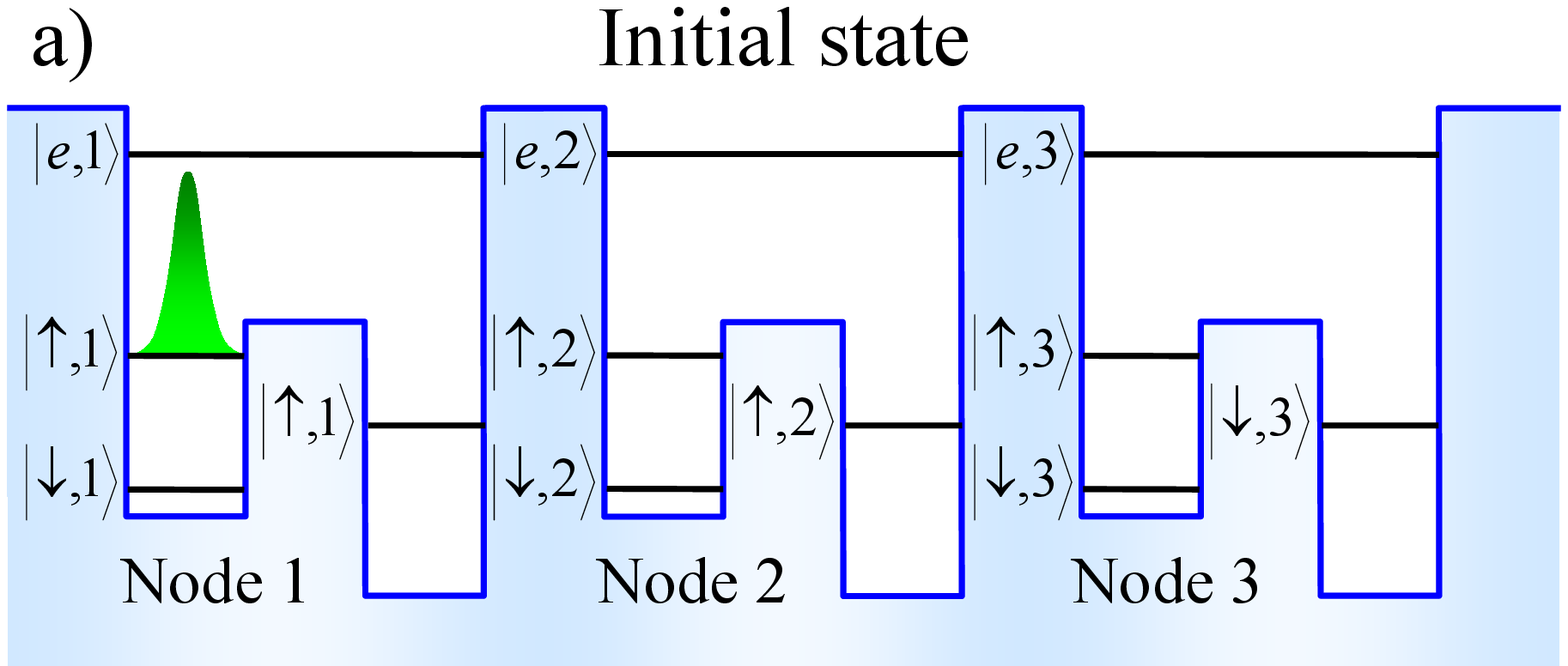}
    \includegraphics[width=8cm, bb=0 0 545 240,clip]{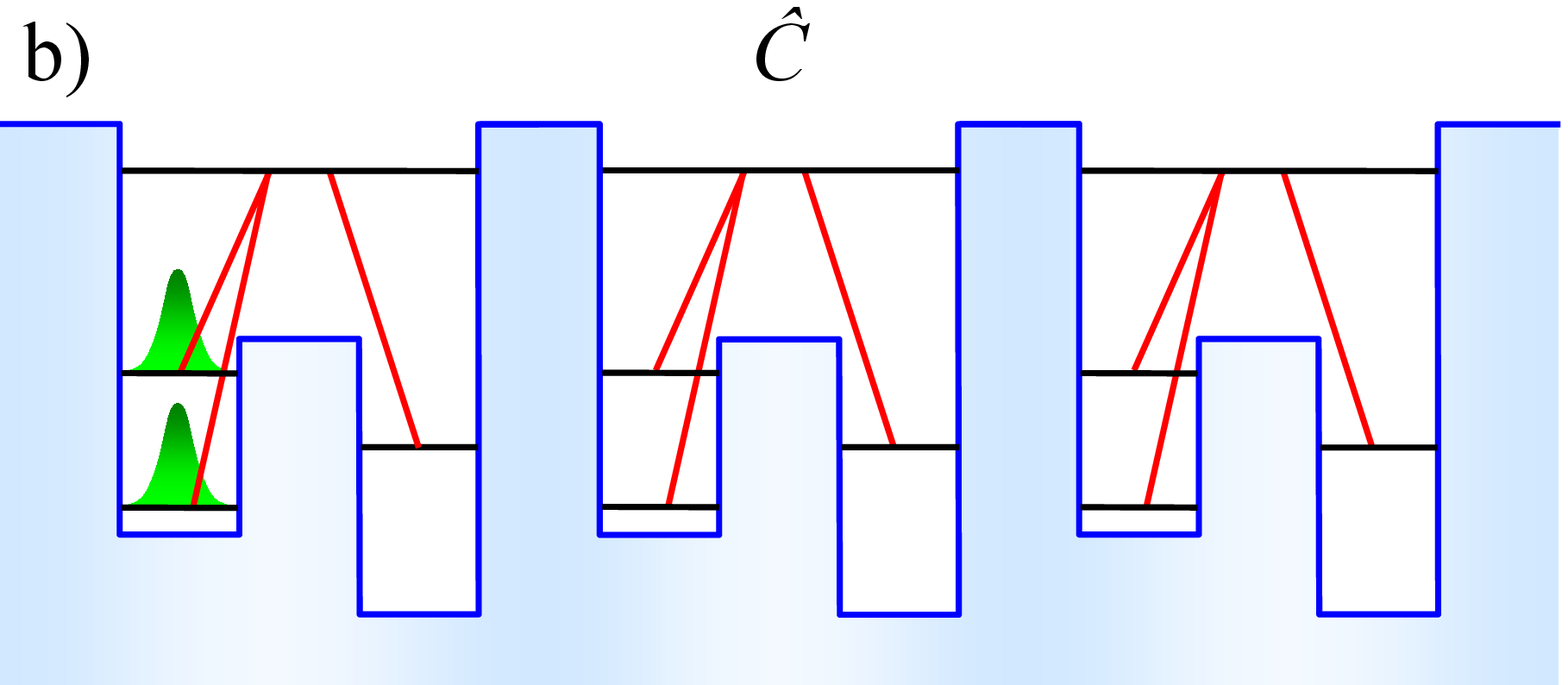}
    \includegraphics[width=8cm, bb=0 0 545 240,clip]{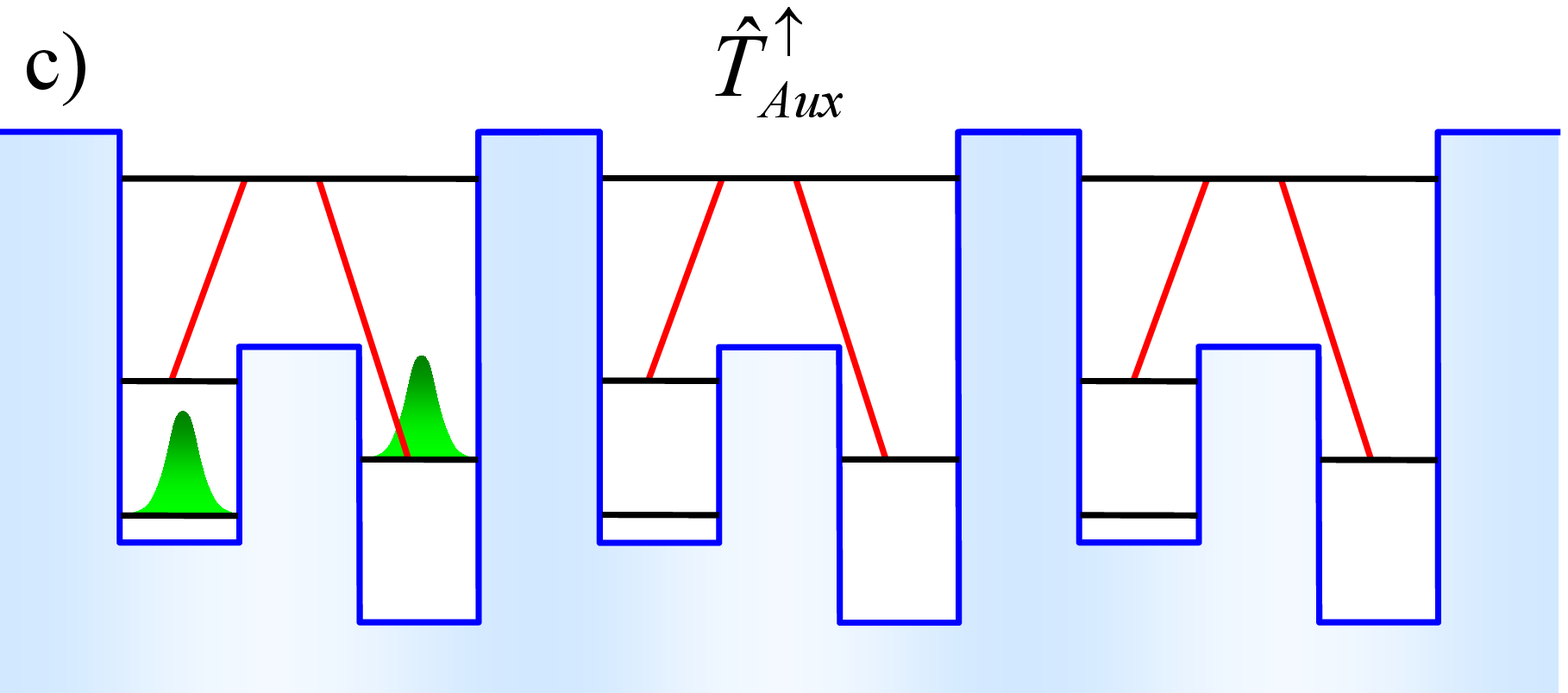}
    \includegraphics[width=8cm, bb=0 0 545 240,clip]{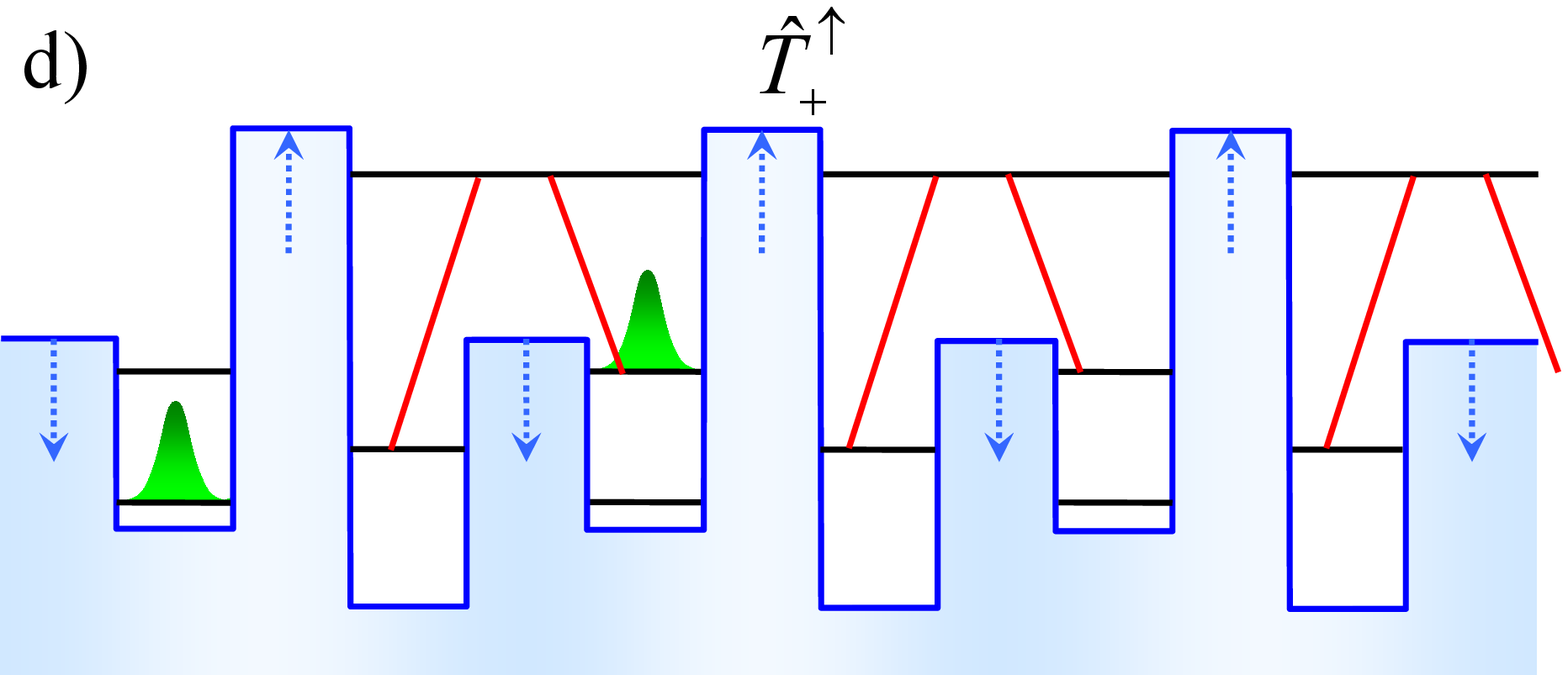}
    \includegraphics[width=8cm, bb=0 0 545 240,clip]{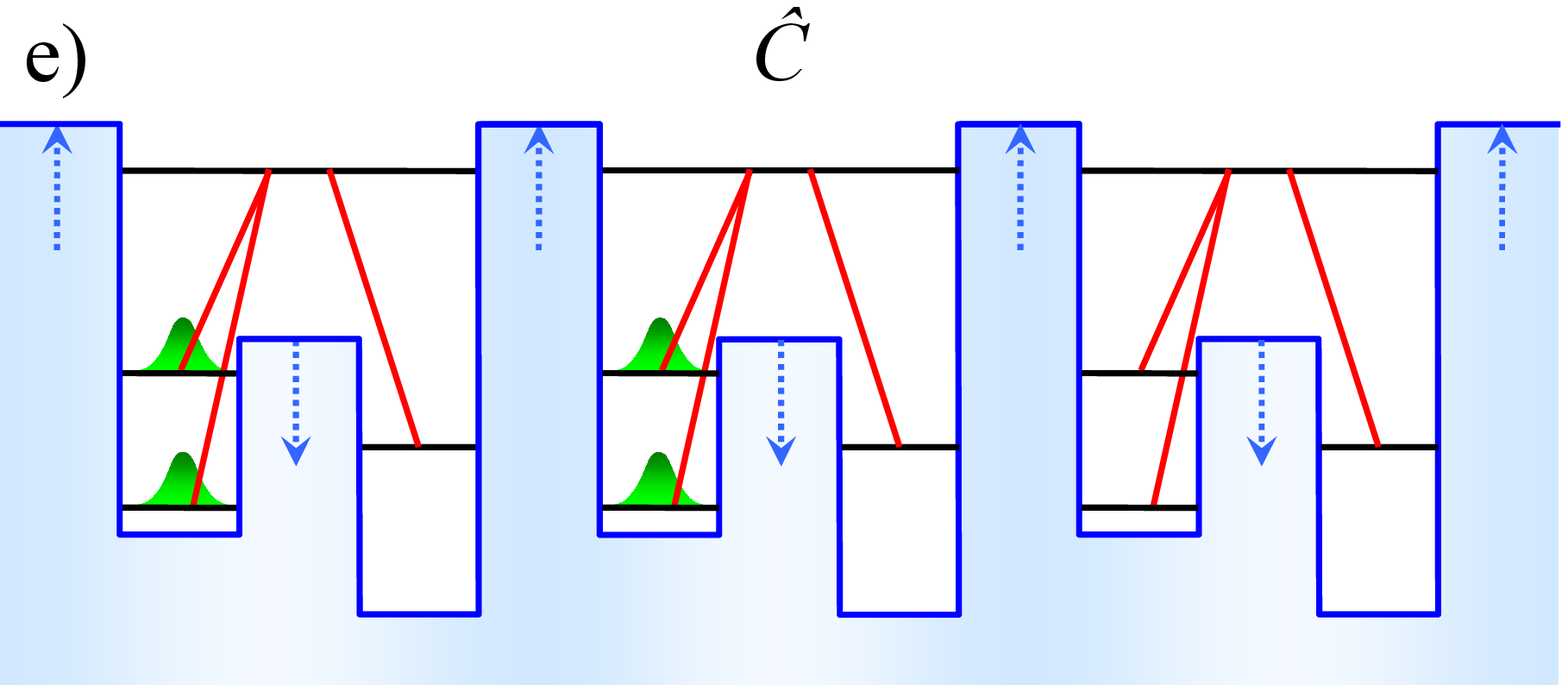}
    \caption{The sequence of 2-photon and 3-photon STIRAP to implement a single step of the quantum walk.
    a) The initial state of the walk with the electron confined to state $\ket{\uparrow,1}$;
    b) A 3-photon STIRAP implements the coin rotation $\hat{C}$, mixing the states $\ket{\uparrow,i}$ and $\ket{\downarrow,i}$;
    c) A 2-photon STIRAP transfers the population from state $\ket{\uparrow,i}$ to state $\ket{A,i}$;
    d) Even barriers are lowered and odd barriers are raised in order to regroup the quantum
    dots. It is now possible for another 2-photon STIRAP to transfers the population from state $\ket{A,i}$ to state
    $\ket{\uparrow,i+1}$, completing the translation operation
    $\hat{T}^{\uparrow}_{+1}$; e) Potential barriers are returned to their initial setting and
    the above process repeated.}
    \label{fig.qrw-procedure}
\end{figure}

\begin{figure}
    \centering
    \includegraphics[width=8cm]{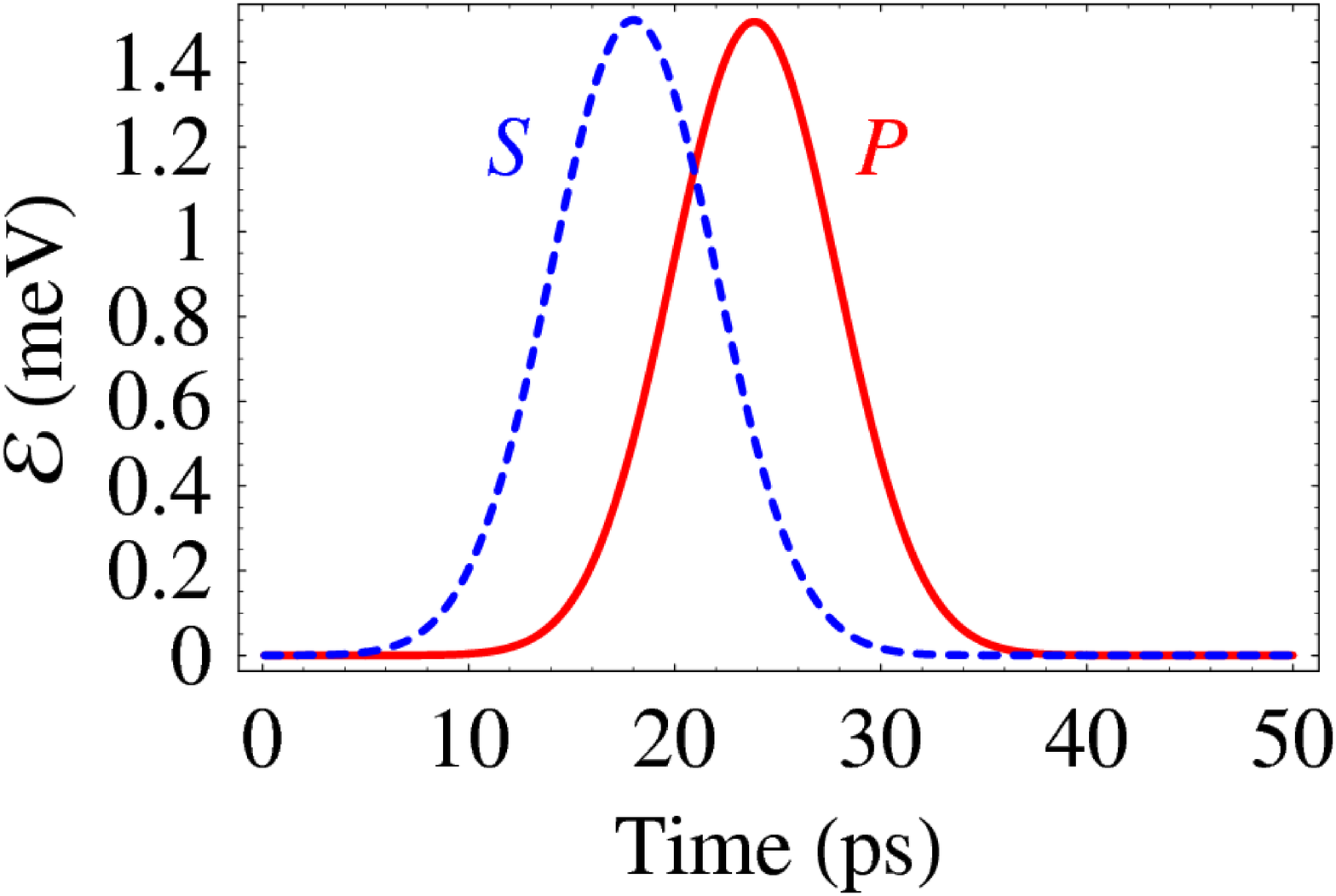}
    \includegraphics[width=9cm]{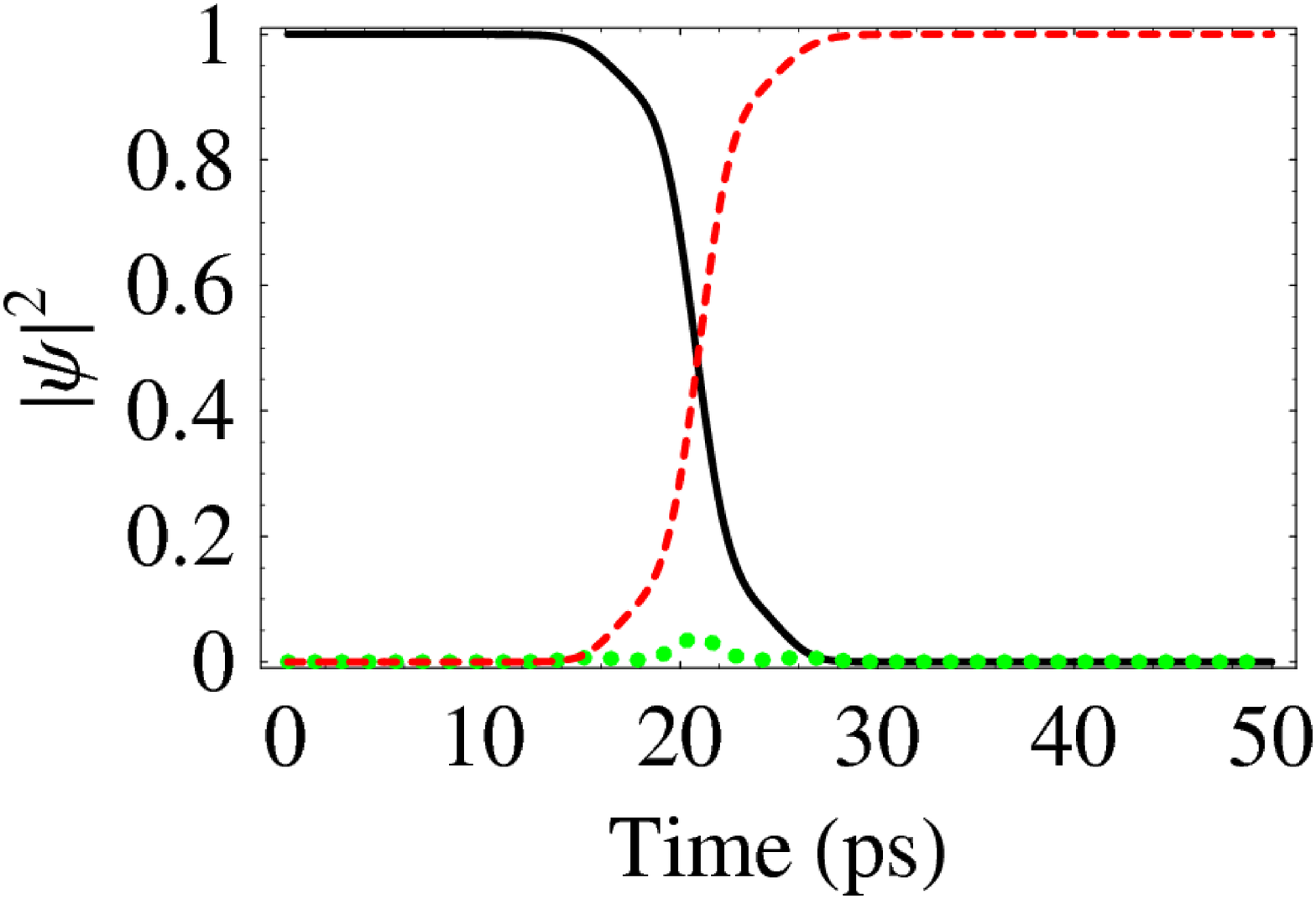}
    \caption{(upper panel) The two laser pulses involved in a 2-photon STIRAP process. Both pulses have a Gaussian envelope
    and are applied in the counter intuitive order, i.e. the Stokes pulse
    $S$ responsible for the $\ket{e}\longleftrightarrow\ket{A}$ transition is applied before
    the pump pulse responsible for the $\ket{\uparrow}\longleftrightarrow\ket{e}$ transition.
    (lower panel) The time evolution of dressed states $\ket{\uparrow}$ (solid), $\ket{e}$
    (dotted) and $\ket{A}$ (dashed) due to the application of the 2-photon STIRAP with pulse parameters optimized to perform a swap operation.
    Initially $\psi_{\uparrow}=1$ and $\psi_A=\psi_e=0$.}
    \label{fig.2ph-stirap}
\end{figure}

An important consideration in the design of the quantum dots is the
ability to perform selective addressing of states which are being
coupled via STIRAP and to avoid all unwanted secondary excitations.
Taking, for example, the $\Omega_{\uparrow}$ pulse which is intended
to couple the energy levels $E_{\uparrow}$ and $E_e$, the quantum
dot energy structure should disallow a secondary excited state, say
$E_{e'} = E_{\downarrow}+\Omega_{\uparrow}\hbar$ to exist, as it
would lead to the unwanted excitation of the $E_{\downarrow}$ level.
Similarly, to avoid leaking the electron out of the dot via ladder
excitations, the energy structure should prevent the coupling of
$E_e$ to an upper energy level $E_e + \Omega_{\uparrow}\hbar$. What
is attractive about our proposal is that experimentally this can be
achieved without resorting to complex profiles for the quantum dot
potential. In fact using simple square wells with dimensions given
in Fig. \ref{fig.qdot-node-scheme}b we were able to produce the
necessary energy structure depicted in Fig.
\ref{fig.qdot-selective-coupling}. Assuming absorption line widths
$\alpha \lesssim 1$ meV, all superfluous excitations will be far off
resonance and will not have any appreciable magnitude.

\section{Results} \label{section.results}

In order to simulate the evolution of the quantum walk in the array
of dots, we first tuned our laser parameters to correctly perform
the desired $\hat{C}$ rotation and $\hat{T}^{\uparrow}_{+1}$
translation operations. For our 2-photon STIRAP operations we employ
pump and Stokes pulses with Gaussian envelopes $\mathcal{E}_{s}(t)$
and $\mathcal{E}_{p}(t)$ and parameterize them using their peak
interaction energies $\mathcal{\overline{E}}_{p}$ and
$\mathcal{\overline{E}}_{s}$, standard deviations $\sigma_{p}$ and
$\sigma_{s}$, phase angles $\alpha_{p}$ and $\alpha_{s}$, and the
time interval $\Delta \mathcal{T}$ between the peak interaction
energies. The STIRAP process can now be modeled by the
time-dependant hamiltonian
\begin{equation}
    \hat{H}(t) = \left(
    \begin{array}{ccc}
        0 & \mathcal{E}_{s}(t)e^{i\alpha_{s}} & 0 \\
        \mathcal{E}_{s}(t)e^{-i\alpha_{s}} & 0 & \mathcal{E}_{p}(t)e^{i\alpha_{p}} \\
        0 & \mathcal{E}_{p}(t)e^{-i\alpha_{p}} & 0
    \end{array} \right),
\end{equation}
constructed using the Rotating Wave Approximation
\cite{book.Shore2006} in the Raman resonance limit. The pulse
parameters need to be tuned such that the resulting 2-photon STIRAP
operation coherently transfers an entire population from state
$\ket{\uparrow, i}$ to state $\ket{A, i}$ via state $\ket{e, i}$. We
achieve this by optimizing $\mathcal{\overline{E}}_{p}$ and $\Delta
\mathcal{T}$, using a technique detailed in the appendix, while
other parameters are fixed to any desired values. Figure
\ref{fig.2ph-stirap} shows the time evolution of the dressed states
$\ket{\uparrow, i}$ and $\ket{A, i}$ and $\ket{e, i}$ under the
application of the optimized 2-photon STIRAP, where we have set
$\mathcal{\overline{E}}_{s}=1.5$ meV, $\sigma_{p}=\sigma_{s}=4.0$ ps
and $\alpha_{p}=\alpha_{s}=0$, and the optimum
$\mathcal{\overline{E}}_{p}=1.50$ meV and $\Delta \mathcal{T} =
5.87$ ps. In order to achieve the second transition from $\ket{A,
i}$ to $\ket{\uparrow, i+1}$ (after raising and lowing the alternate
potential barriers) we simply reverse the order in which the pump
and Stoke pulses are applied.

\begin{figure}
    \centering
    \includegraphics[width=8cm]{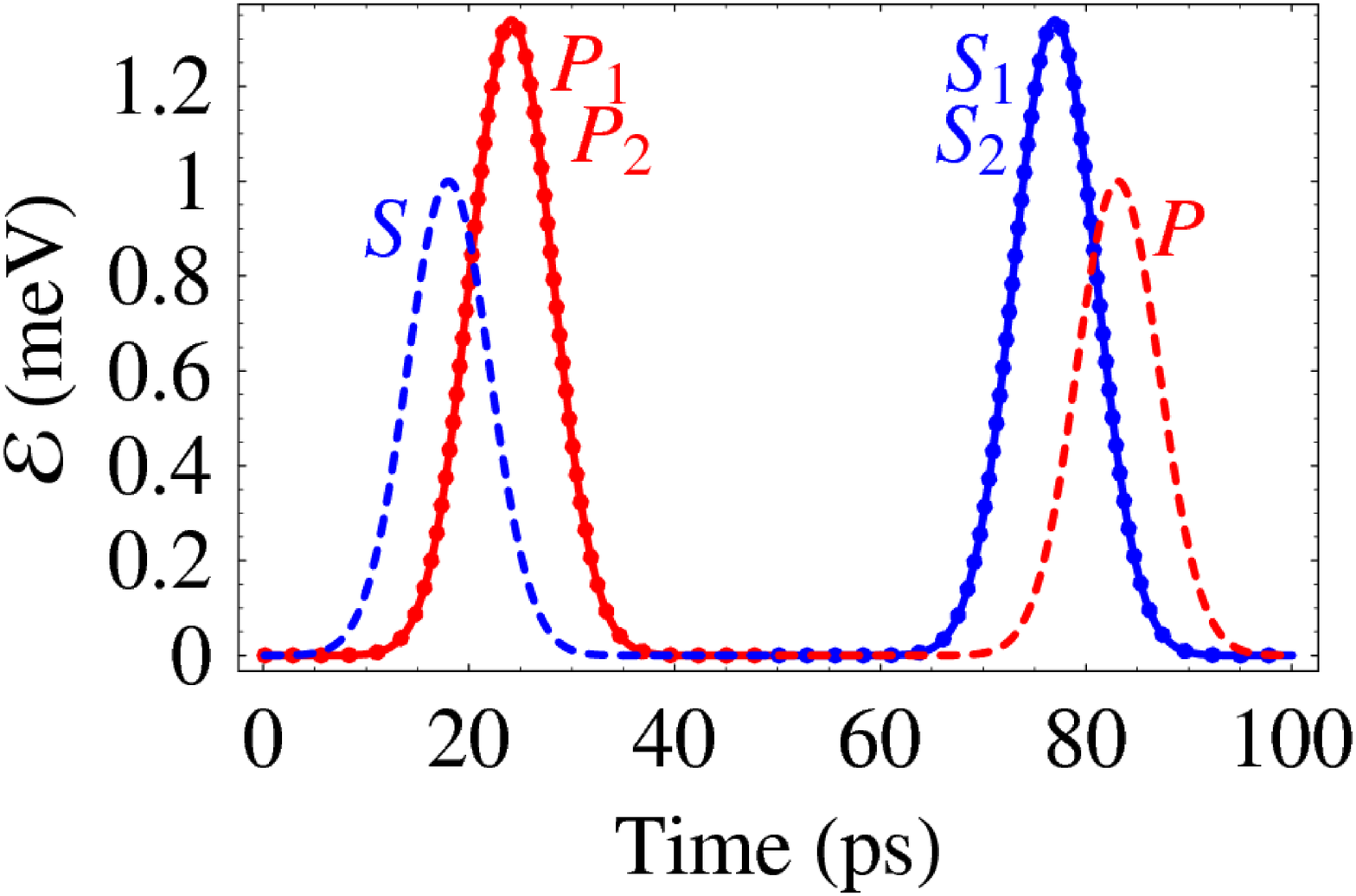}
    \includegraphics[width=9cm]{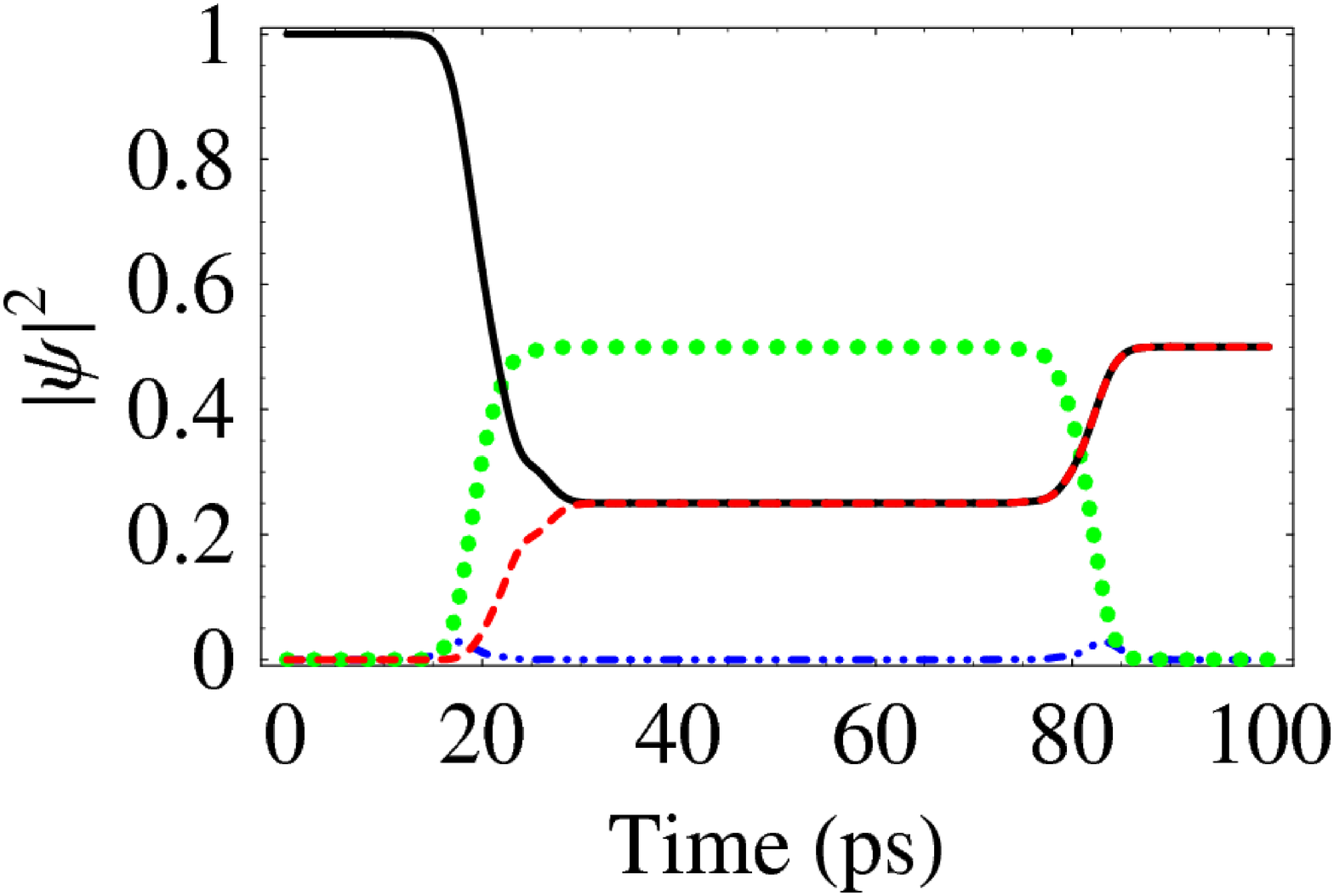}
    \includegraphics[width=9cm]{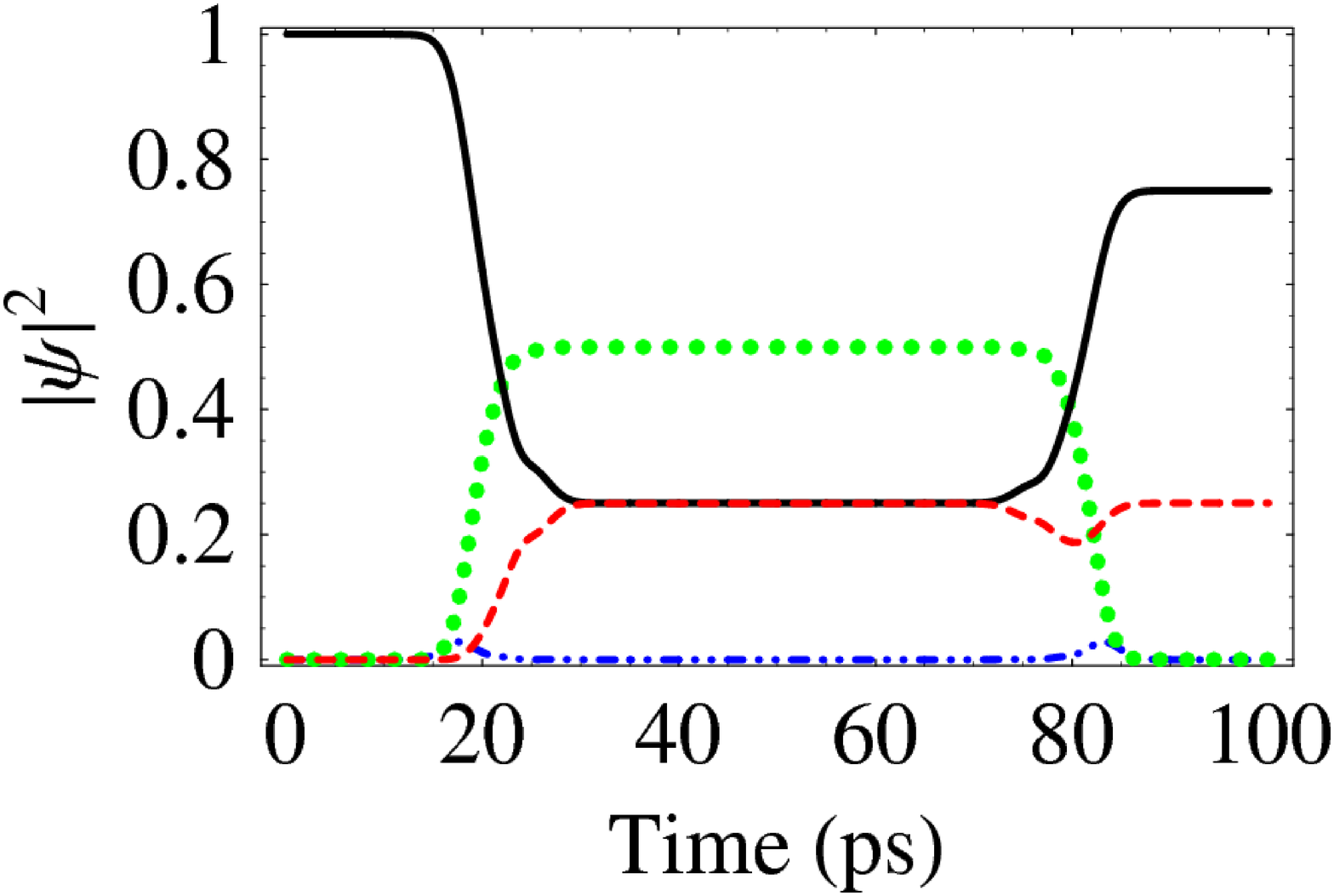}
    \caption{(upper panel) The six laser pulses involved in a 3-photon STIRAP process. All pulses have a Gaussian envelope
    and are applied in the counter intuitive order.
    (lower panels) The time evolution of dressed states $\ket{\uparrow}$ (solid), $\ket{\downarrow}$
    (dashed), $\ket{e}$ (dot-dashed) and $\ket{A}$ (dotted) with pulse parameters optimized to implement the coin
    operators $\hat{C}(\nicefrac{\pi}{4}, \nicefrac{\pi}{2}, \nicefrac{\pi}{2})$
    and $\hat{C}(\pi/6, \nicefrac{\pi}{2}, \nicefrac{\pi}{2})$ respectively.
    Initially $\psi_{\uparrow}=1$ and $\psi_{\downarrow}=\psi_A=\psi_e=0$.}
    \label{fig.3ph-stirap}
\end{figure}

In the double 3-photon STIRAP process depicted in Fig.
\ref{fig.3ph-stirap}, the pulse parameters need to be tuned to
perform a unitary operation $\hat{C}$ on the coin states. As before
we achieve this by optimizing $\mathcal{\overline{E}}_{b}$ and
$\Delta \mathcal{T}$ after fixing the other parameters to any
desired values. Setting $\mathcal{\overline{E}}_{a}=1.0$ meV,
$\sigma=4.0$ ps, $\alpha_{s}=\alpha_{p1}=\beta_{p}=0$,
$\alpha_{p2}=\beta_{s1}=\beta_{s2}=\pi$,
$\alpha_{p1}=\beta_{s1}=\pi$, $\alpha_{p2}=\beta_{s2}=0$ and
$\alpha_s=0$, with optimum parameters
$\mathcal{\overline{E}}_{b}=1.34$ meV and $\Delta \mathcal{T}=6.12$
ps, we obtain symmetric coins by simply varying $\beta_p$. Figure
\ref{fig.3ph-stirap} depict the time evolution of the dressed states
$\ket{\downarrow}$, $\ket{\uparrow}$, $\ket{A}$ and $\ket{e}$ under
the action of the coin operators $\hat{C}(\nicefrac{\pi}{4},
\nicefrac{\pi}{2}, \nicefrac{\pi}{2})$ and $\hat{C}(\pi/6,
\nicefrac{\pi}{2}, \nicefrac{\pi}{2})$ for
$\beta_p=\nicefrac{\pi}{2}$ and $\beta_p=\nicefrac{\pi}{3}$
respectively. We also obtain asymmetric coins like
$\hat{C}(\nicefrac{\pi}{4}, \nicefrac{\pi}{2}, -\nicefrac{\pi}{2})$
by setting $\alpha_{p1}=\beta_{s1}=\pi$,
$\alpha_{p2}=\beta_{s2}=\nicefrac{\pi}{2}$, $\alpha_s=0$ and
$\beta_p=\nicefrac{\pi}{2}$.

\begin{figure}
    \centering
    \includegraphics[width=8cm]{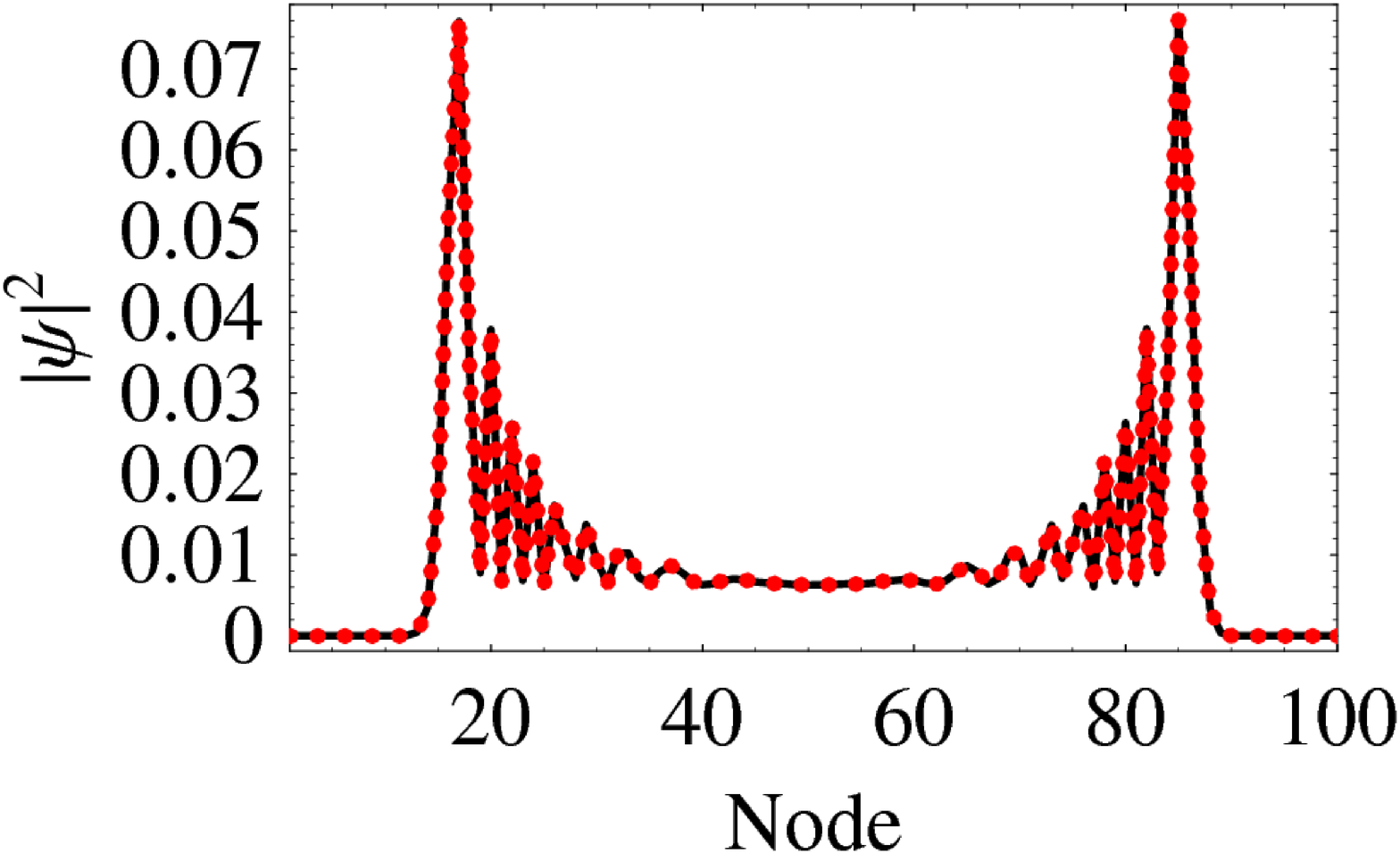}
    \includegraphics[width=8cm]{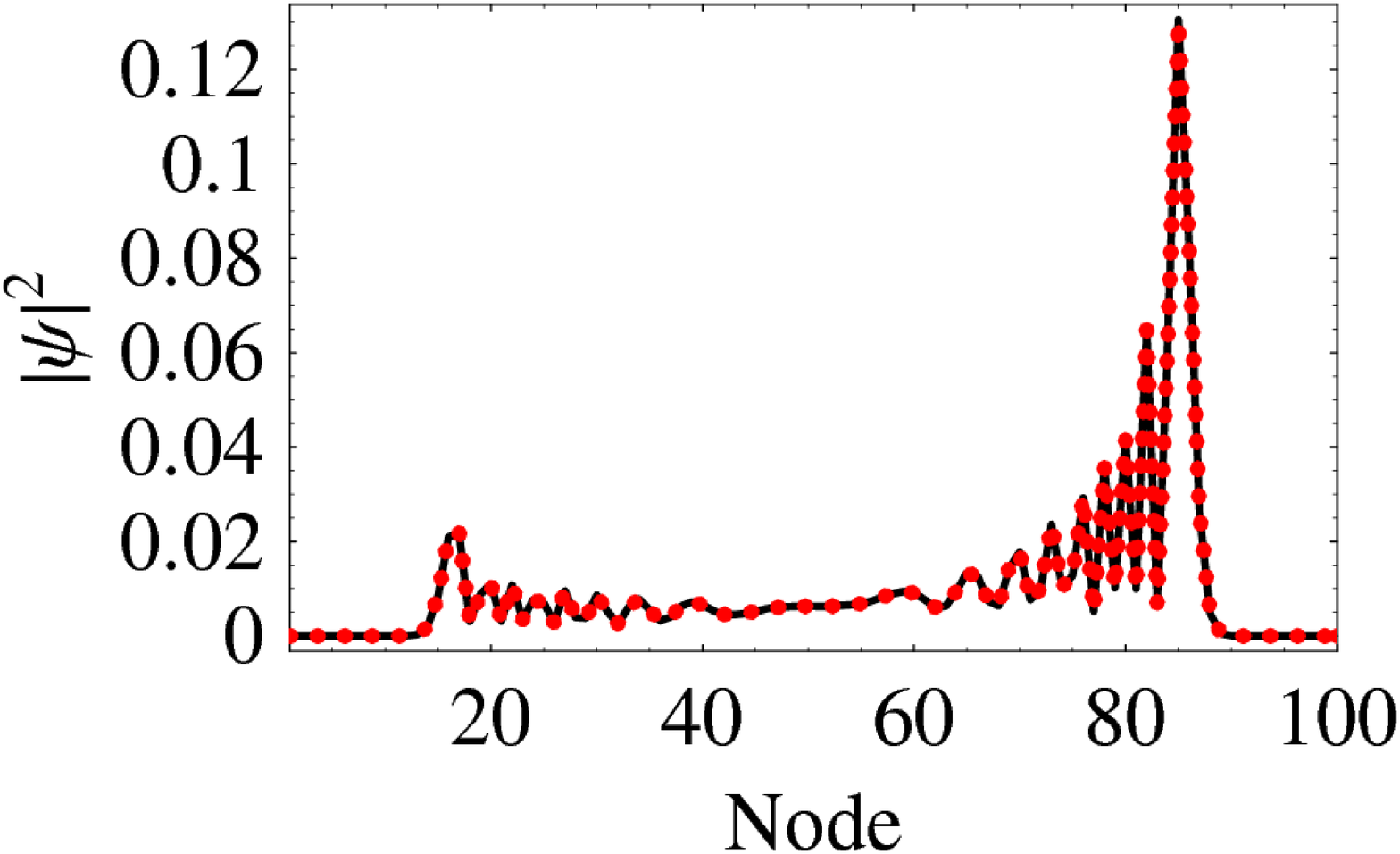}
    \caption{The electron probability distribution after
    100 applications of the pulse sequence (solid) vs. the corresponding discrete-time quantum random walk
    distribution after 100 steps (dotted). The pulse 3-photon pulse
    parameters were optimized to perform $\hat{C}(\pi/4, \nicefrac{\pi}{2},
    \nicefrac{\pi}{2})$ and the electron was initially confined to node 1 with probability distribution:
    $\psi_{\uparrow}=\psi_{\downarrow}=\nicefrac{1}{\sqrt{2}}$ (upper panel) and $\psi_{\uparrow}=1$ and
    $\psi_{\downarrow}=0$ (lower panel).}
    \label{fig.qrw-result}
\end{figure}

\begin{figure}
    \centering
    \includegraphics[width=7cm]{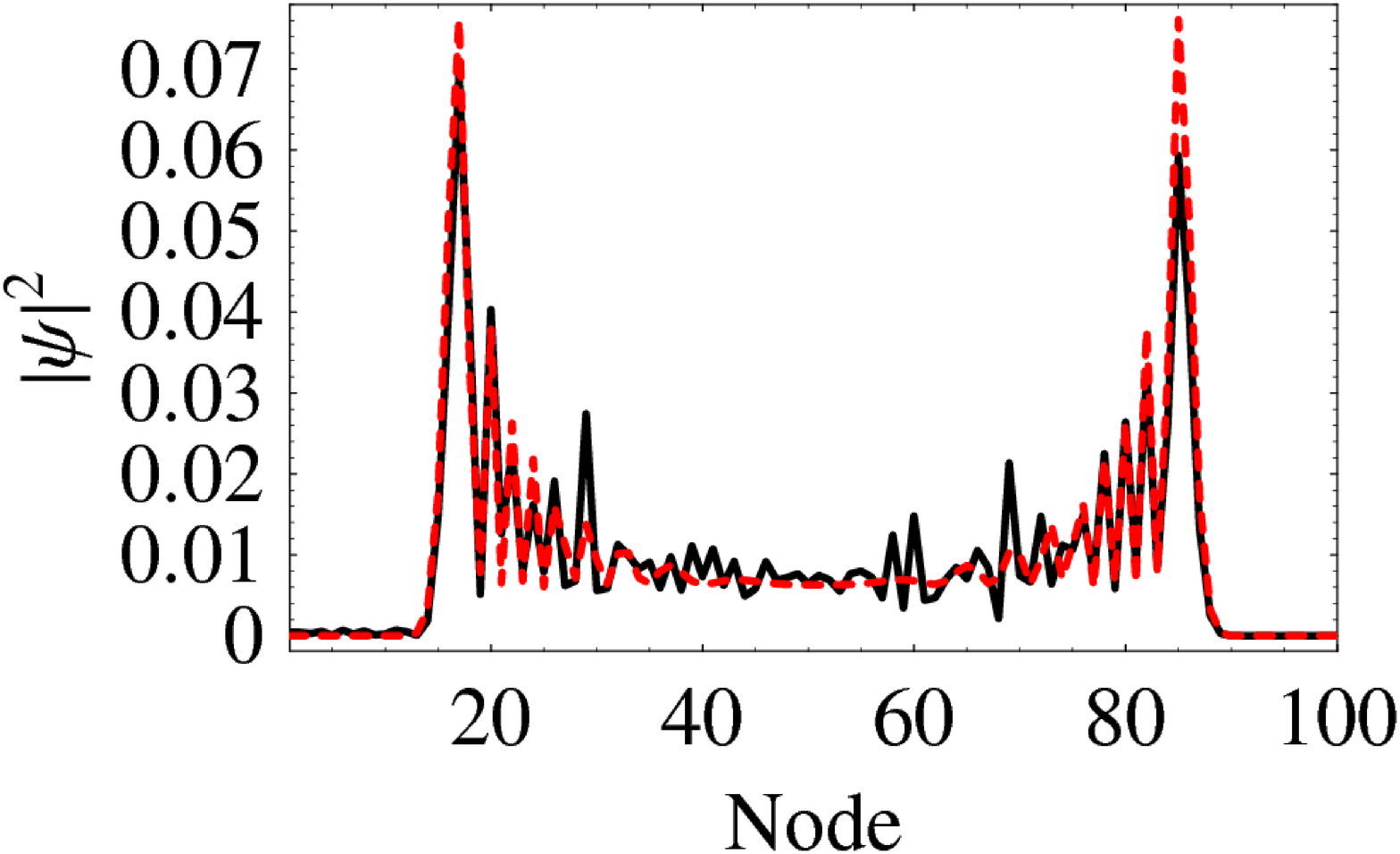}
    \includegraphics[width=7cm]{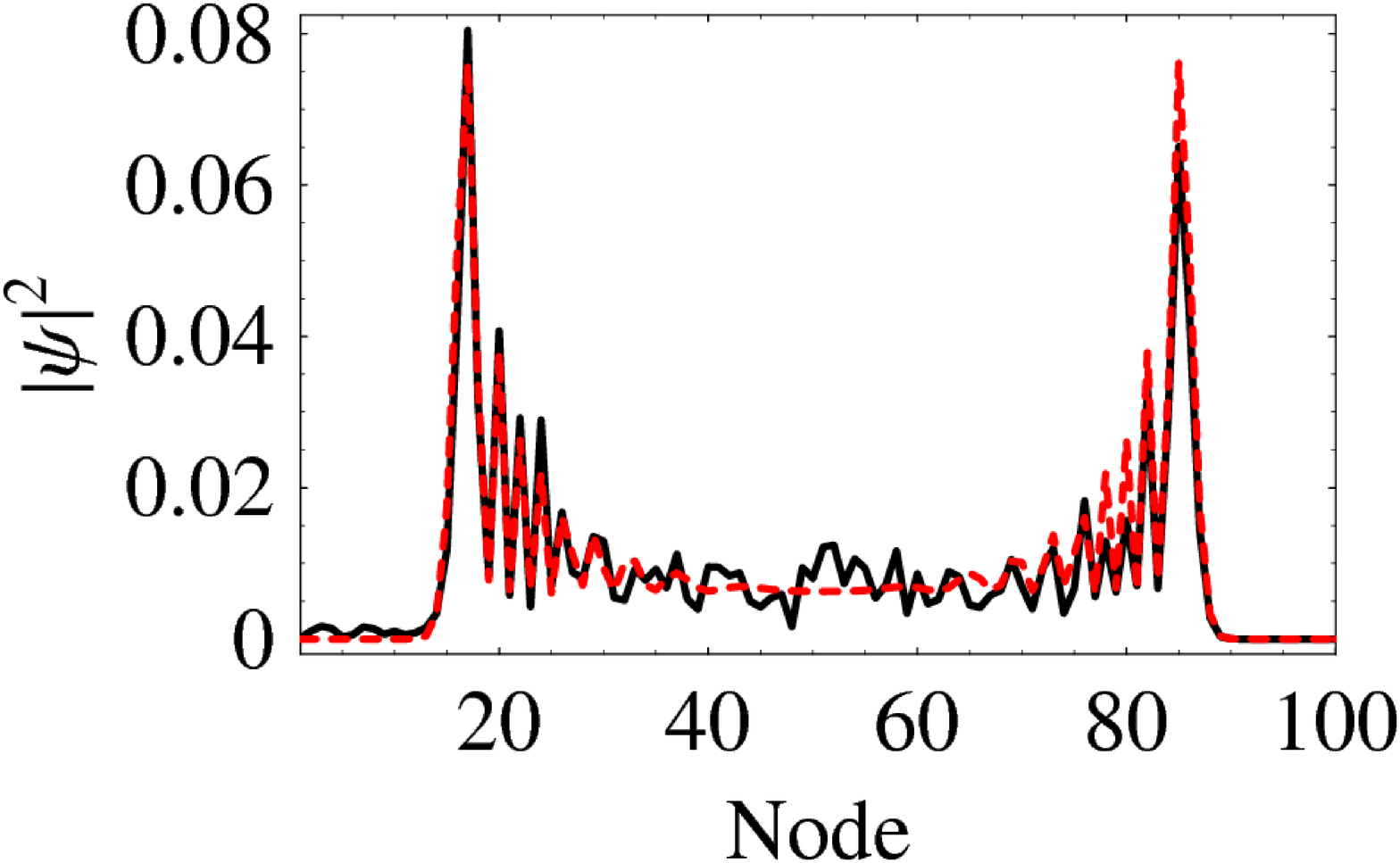}
    \includegraphics[width=7cm]{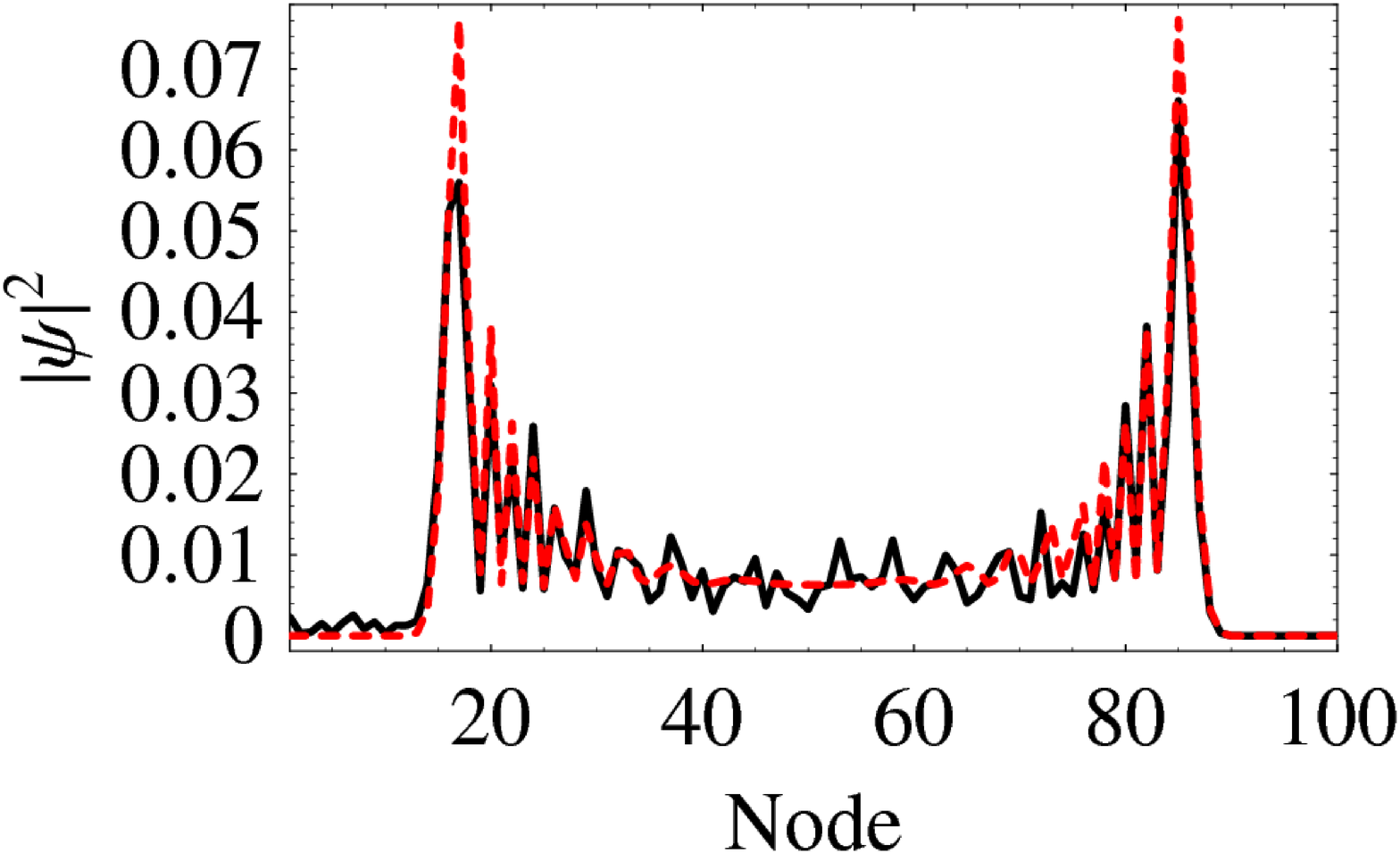}
    \includegraphics[width=7cm]{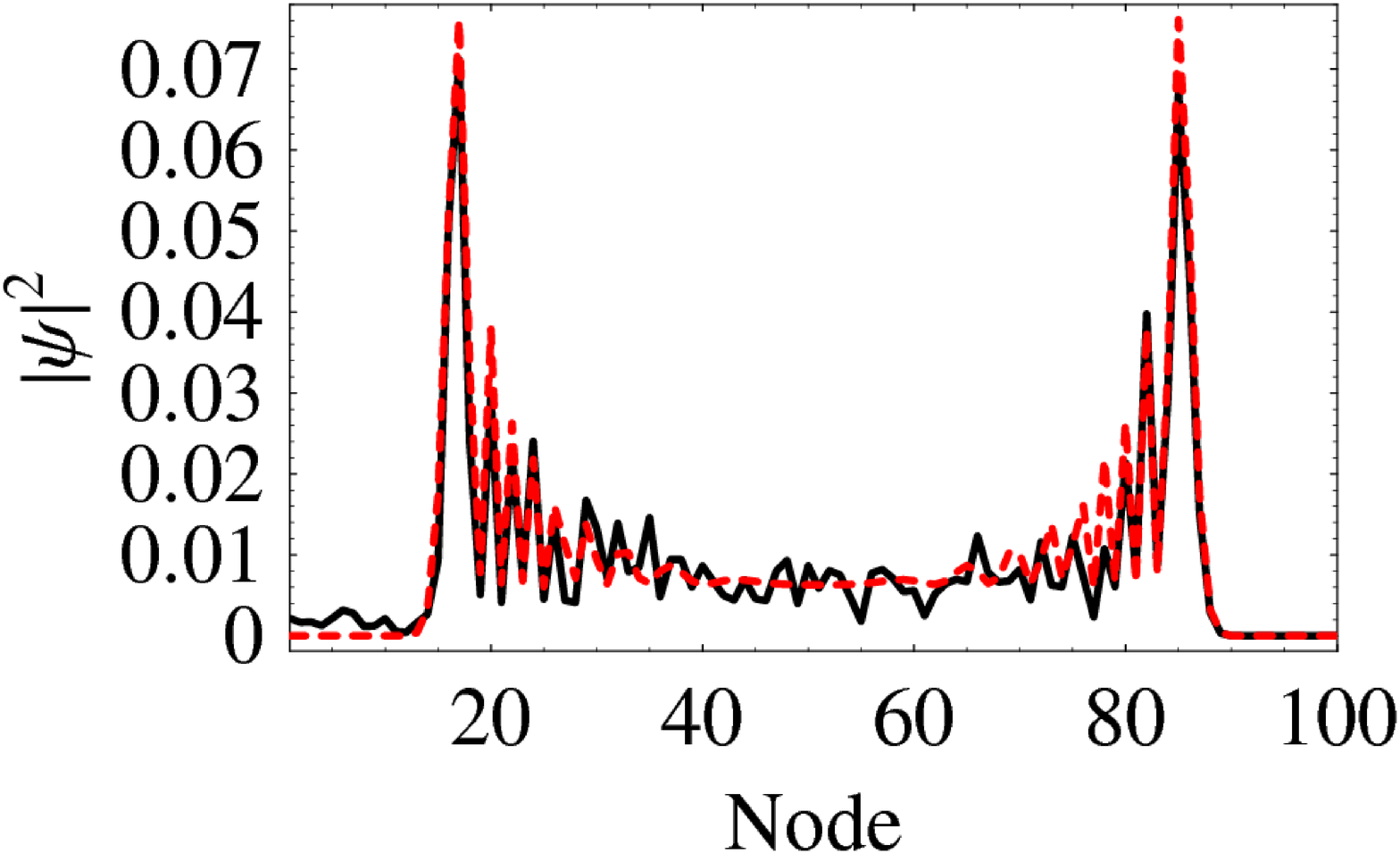}
    \caption{The impact of introducing some noise in the applied
    pulse parameters.
    a) Deviation from the exact quantum random walk
    distribution (dashed) due to an induced 2\% uncertainty in the laser pulse peak energies (solid).
    b) Deviation from the exact quantum random walk
    distribution (dashed) due to an induced 5\% uncertainty in the laser pulse phases (solid).
    c) Deviation from the exact quantum random walk
    distribution (dashed) due to an induced 2\% uncertainty in the laser pulse standard deviations (solid).
    d) Deviation from the exact quantum random walk
    distribution (dashed) due to an induced 0.3\% uncertainty in the laser pulse timing (solid).}
    \label{fig.qrw-error}
\end{figure}

Following the control pulse optimization, we obtain the full
$3\times3$ and $4\times4$ evolution matrices corresponding to the 2-
and 3-photon STIRAP operations respectively. We then use these to
simulate the evolution of a single electron under the repeated
applications of the pulse sequence outlined in Fig.
\ref{fig.qrw-procedure}. In Fig. \ref{fig.qrw-result} we have
plotted the electron wavefunction after 100 steps, using optimized
pulses corresponding to the translation operator
$\hat{T}^{\uparrow}_{+1}$ as well as two different coin operators
$\hat{C}(\nicefrac{\pi}{4}, \nicefrac{\pi}{2}, \nicefrac{\pi}{2})$,
and $\hat{C}(\nicefrac{\pi}{6}, \nicefrac{\pi}{2},
\nicefrac{\pi}{2})$. The results are in excellent agreement with
their respective ideal theoretical distributions. We also
investigated the effect of noise disturbance and experimental
uncertainty on the resulting distribution and demonstrated a
relatively robust response against imperfect pulse parameters.
Figure \ref{fig.qrw-error} shows a reasonable degree of fidelity
after the introduction of white noise in the energy peak, phase,
timing and the standard deviation of the laser pulses.

\section{Conclusion}

We have proposed a physical implementation of a discrete-time
quantum random walk using the action of 2- and 3-photon STIRAP
operations on an array of quantum dots. We demonstrated that our
scheme reproduces the characteristic quantum walk probability
distribution which remains observable after the introduction of
modest experimental uncertainty in the laser excitations.

Like many other proposed schemes however, our implementation of the
quantum walk is essentially a wave interference experiment and does
not involve any quantum entanglements. Such implementations come
with a cost as the number of resources grows, at best linearly with
the number of nodes required for the walk. Furthermore it is
generally expected that almost all potentially useful applications
of quantum walks such as search algorithms \cite{Shenvi2003} or
element distinctness \cite{Ambainis2003}, stem from higher
dimensional walks on general graphs. Nevertheless implementing one
dimensional quantum walks is significant for carrying out
feasibility studies of assembling such physical systems.

\newpage

\renewcommand{\theequation}{A-\arabic{equation}}
\setcounter{equation}{0}
\section*{APPENDIX: Control Pulse Optimization}
\label{appendix.pulse-optimization}

Considering a time-dependant hamiltonian $\hat{H}(t)$ for the
2-photon STIRAP, its action on a three-level system can be
determined by solving the Schr\"{o}dinger equation
\begin{equation}
    \hat{H} \left(
    \begin{array}{c}
        \psi_{1}(t) \\
        \psi_{2}(t) \\
        \psi_{3}(t)
    \end{array} \right) =
    i\hbar \frac{\partial}{\partial t}\left(
    \begin{array}{c}
        \psi_{1}(0) \\
        \psi_{2}(0) \\
        \psi_{3}(0)
    \end{array} \right).
\end{equation}
We do this by approximating $\hat{H}(t)$ using a series of time
independent $\hat{H}_{i}$ over suitably short time steps $\delta t$,
which allows us to write the solution as
\begin{equation}
    \left(
    \begin{array}{c}
        \psi_{1}(t) \\
        \psi_{2}(t) \\
        \psi_{3}(t)
    \end{array} \right) =
    \hat{U}_T(t) \left(
    \begin{array}{c}
        \psi_{1}(0) \\
        \psi_{2}(0) \\
        \psi_{3}(0)
    \end{array} \right),
\end{equation}
where the evolution operator
\begin{eqnarray}
    \hat{U}_{T}(t) & = &
    e^{-i\hat{H}_{t}\delta t / \hbar}
    e^{-i\hat{H}_{t-1}\delta t / \hbar}
    \cdots
    e^{-i\hat{H}_{2}\delta t / \hbar}
    e^{-i\hat{H}_{1}\delta t / \hbar} \\
    & = & \left(
    \begin{array}{ccc}
        \hat{u}_{11} & \hat{u}_{12} & \hat{u}_{13} \\
        \hat{u}_{21} & \hat{u}_{22} & \hat{u}_{23} \\
        \hat{u}_{31} & \hat{u}_{32} & \hat{u}_{33}
    \end{array} \right).
\end{eqnarray}
By carefully optimizing the pulse parameters we can achieve
\begin{equation}
    \left(
    \begin{array}{cc}
        \hat{u}_{11} & \hat{u}_{13} \\
        \hat{u}_{31} & \hat{u}_{33} \\
    \end{array} \right)
    \simeq \hat{T} =
    \left(
    \begin{array}{cc}
        0 & 1 \\
        1 & 0 \\
    \end{array} \right),
    \label{eqn-trans-opt}
\end{equation}
which is the ideal swap operation between states $\ket{1}$ and
$\ket{3}$ via the intermediate state $\ket{2}$. When state $\ket{3}$
is initially empty, this amounts to a translation operation which
coherently transfers an amplitude from state $\ket{1}$ entirely to
the empty state $\ket{3}$ without populating the intermediate state
$\ket{2}$.

We achieve the optimization by first fixing
$\mathcal{\overline{E}}_{s}$ and phase angles $\sigma_{p}$,
$\sigma_{s}$, $\alpha_{p}$ and $\alpha_{s}$, and then varying
$\mathcal{\overline{E}}_{p}$ and $\Delta \mathcal{T}$ in order to
minimize the cost function
\begin{eqnarray}
    \kappa_{T} &=& \sum\abs{
    \left(
    \begin{array}{cc}
        \hat{u}_{11} & \hat{u}_{13} \\
        \hat{u}_{31} & \hat{u}_{33} \\
    \end{array} \right)-
    \left(
    \begin{array}{cc}
        0 & 1 \\
        1 & 0 \\
    \end{array} \right)} \\
    &=& \abs{\hat{u}_{11}} + \abs{\hat{u}_{13} - 1} +
        \abs{\hat{u}_{31} - 1} + \abs{\hat{u}_{33}}.
    \label{eqn-opt-costfunc-trans}
\end{eqnarray}

The exponentials $e^{-i\hat{H}_{t}\delta t / \hbar}$, which have to
be re-evaluated for every parameter variation, are efficiently and
accurately computed using a Chebyshev expansion \cite{Tal-Ezer1984,
Wang1999}
\begin{equation}
    e^A=\sum_{n=0}^\mathcal{N}a_n(\alpha)\phi_n(\mathcal{\widetilde{A}}),
\end{equation}
where $A=-i\hat{H}_{t}\delta t / \hbar$, $a_n(\alpha)= 2J_n(\alpha)$
except for $a_0(\alpha)=J_0(\alpha)$, $J_n(\alpha)$ are the Bessel
functions of the first kind, $\phi_n$ are the Chebyshev polynomials,
and $\mathcal{N}$ is the number of terms in the Chebyshev expansion.
To ensure convergence, the exponent $A$ needs to be normalized as
\begin{equation}\label{eqn.Chebyshev}
    \mathcal{\widetilde{A}}=\frac{2A}{\mu_{max}-\mu_{min}},
\end{equation}
where $\mu_{min}$ and $\mu_{max}$ represent the minimum and maximum
eigenvalues of $A$. Chebyshev polynomials $\phi_n$ are efficiently
evaluated using the recurrence relation
\begin{equation}\label{eqn.Chebyshev}
    \phi_n(\mathcal{\widetilde{A}})=2\mathcal{\widetilde{A}}\phi_{n-1}(\mathcal{\widetilde{A}})+\phi_{n-2}(\mathcal{\widetilde{A}}),
\end{equation}
and
\begin{equation}\label{eqn.Chebyshev}
    \phi_0(\mathcal{\widetilde{A}})=1, \text{~~}
    \phi_1(\mathcal{\widetilde{A}})=\mathcal{\widetilde{A}}.
\end{equation}
In practice, iterations are continued until the norm of the matrix
exponential converges to the required level of accuracy.

The 3-photon STIRAP process is similarly represented by
\begin{equation}
    \left(
    \begin{array}{c}
        \psi_{1}(t) \\
        \psi_{2}(t) \\
        \psi_{3}(t) \\
        \psi_{4}(t)
    \end{array} \right) =
    \hat{U}_{C}(t) \left(
    \begin{array}{c}
        \psi_{1}(0) \\
        \psi_{2}(0) \\
        \psi_{3}(0) \\
        \psi_{4}(0)
    \end{array} \right),
\end{equation}
where the evolution operator
\begin{equation}
    \hat{U}_{C}(t) = \left(
    \begin{array}{cccc}
        \hat{u}_{11} & \hat{u}_{12} & \hat{u}_{13} & \hat{u}_{14}\\
        \hat{u}_{21} & \hat{u}_{22} & \hat{u}_{23} & \hat{u}_{24}\\
        \hat{u}_{31} & \hat{u}_{32} & \hat{u}_{33} & \hat{u}_{34}\\
        \hat{u}_{41} & \hat{u}_{42} & \hat{u}_{43} & \hat{u}_{44}
    \end{array} \right).
\end{equation}
This time pulse parameters can be optimized in order to to achieve
\begin{equation}
    \left(
    \begin{array}{cc}
        \hat{u}_{11} & \hat{u}_{12} \\
        \hat{u}_{21} & \hat{u}_{22} \\
    \end{array} \right)\simeq \hat{C},
\end{equation}
where $\hat{C}$ is a desired unitary coin matrix given by Eq.
\ref{eqn.unitary-coin}.

\begin{figure}
    \centering
    \includegraphics[width=12cm]{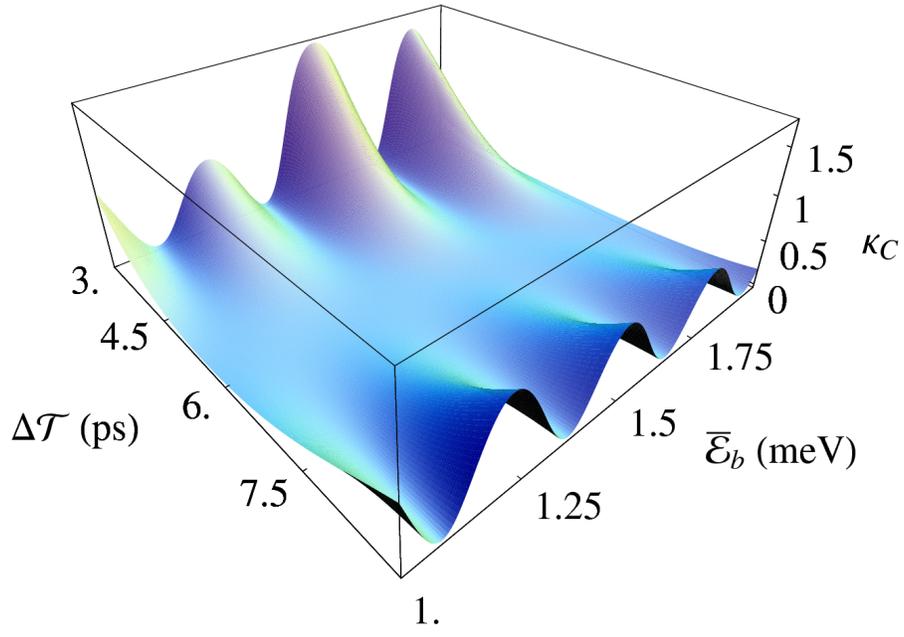}
    \caption{The optimization surface of the $\hat{C}$ operator obtained by minimizing the cost function $\kappa_{C}$
    (Eq. \ref{eqn-opt-costfunc-coin}). Referring to Fig. \ref{fig.3ph-stirap}, $\mathcal{\overline{E}}_b$ is the $P1$ pulse energy peak
    and $\Delta \mathcal{T}$ is the time between the $S$ and $P1$ pulse energy peaks.}
    \label{fig.coin-opt-surface}
\end{figure}

As before, this is achieved by fixing all the parameters except for
$\mathcal{\overline{E}}_{b}$ and $\Delta \mathcal{T}$ which are
varied to minimize the cost function
\begin{equation}
    \kappa_{C} = \sum \abs{
    \left(
    \begin{array}{cc}
        \hat{u}_{11} & \hat{u}_{12} \\
        \hat{u}_{21} & \hat{u}_{22} \\
    \end{array} \right)
    \left(
    \begin{array}{cc}
        \hat{u}_{11} & \hat{u}_{21} \\
        \hat{u}_{12} & \hat{u}_{22} \\
    \end{array} \right)^\ast -
    \left(
    \begin{array}{cc}
        0 & 1 \\
        1 & 0 \\
    \end{array} \right)}.
    \label{eqn-opt-costfunc-coin}
\end{equation}
Figure \ref{fig.coin-opt-surface} shows the minimization surface
profile for the parameters given in Sec. \ref{section.results}. It
is important to note that the above cost function leads to a
``loose'' optimization in the sense that it does not strictly
optimize the STIRAP into any specific coin operator. Rather, it only
requires that the coin matrix be unitary. It also turns out that the
optimum parameters for a unitary $\hat{C}$ are independent of the
choice of phase factors $\alpha$ and $\beta$. Instead these phases
can be conveniently altered to manipulate the exact form of the
operator $\hat{C}$ while maintaining its unitarity.

\newpage

%===========================================
% BIBLIOGRAPHY
%===========================================
\bibliography{qrw-simulation}

\newpage

\section{Figures}

\clearpage

\clearpage

\clearpage

\end{document}